\documentclass[12pt]{article}
\usepackage[authoryear]{natbib}
\usepackage[utf8]{inputenc}
\usepackage[T1]{fontenc}
\usepackage{amsmath, amssymb}
\usepackage{graphicx}
\usepackage{booktabs}
\usepackage[margin=1in]{geometry}
\usepackage{hyperref}
\usepackage{xcolor}
\usepackage{setspace}
\usepackage{caption}
\usepackage{subcaption}
\usepackage{float}

\title{Correcting socioeconomic bias\\in mobile phone mobility estimates\\using multilevel regression and poststratification}

\author{
  Leo Ferres\thanks{Institute of Data Science, Faculty of Engineering, Universidad del Desarrollo, Santiago, Chile; ISI Foundation, Turin, Italy} \and
  L\ae titia Gauvin\thanks{ IRD, UMR 215 Prodig, 5, course of Humanities, F-93 322, Aubervilliers Cedex, France; ISI Foundation, Turin, Italy}
}

\date{}

\begin{document}

\maketitle

\begin{abstract}
Call detail records (CDR) from mobile phone networks are widely used to study human mobility however CDR data from a single mobile operator are inherently biased because the observed users do not mirror the population distribution. Using data from a major Chilean carrier in Santiago, we observe the user base is skewed by socioeconomic group, so aggregate metrics like radius of gyration are distorted by the population that is actually observed.

To correct this sampling bias, we apply multilevel regression and poststratification (MRP), a method that is not yet standard for CDR-based mobility studies. We fit a Bayesian multilevel model for individual mobility using socioeconomic status, gender, and geography, with partial pooling across comunas, and then poststratify the predictions to match census demographics. This approach reduces the naive CDR estimate of average radius of gyration by about 17\%.

Importantly, a version of the model that uses only geographic information still captures much of the bias, showing that MRP can be useful even when the socioeconomic composition of users is not fully known, as long as spatial patterns of socioeconomic groups exist. This example demonstrates how MRP can provide a principled correction for non-representative CDR-derived mobility estimates, rather than treating the carrier sample as if it were a random population sample.

\end{abstract}


\section{Introduction}
\label{sec:intro}

Mobile phone data, particularly Call Detail Records (CDR), have transformed
the study of human mobility \cite{gonzalez2008, barbosa2018}. Passive
observation of millions of users over extended periods enables measurement of
mobility patterns at unprecedented spatial and temporal resolution, with
applications ranging from urban planning \cite{lenormand2015} to epidemic
modelling \cite{wesolowski2012, tizzoni2014,Gozzi2021} and poverty estimation
\cite{blumenstock2015}.

Despite these advances, a fundamental limitation of CDR data is that the user
base of any single mobile carrier is not necessarily a random sample of the population
\cite{ricciato2020, vanhoof2018}. Prior work has documented this primarily
in low-penetration settings, where mobile phone \emph{ownership} itself is
socioeconomically stratified: phone owners in Rwanda and Kenya are wealthier,
better educated, and more predominantly male than the general population
\cite{blumenstock2010, wesolowski2013}, and resulting biases propagate into
modelled outbreak dynamics \cite{chin2025}. However, in high-penetration
markets such as Chile, where mobile subscriptions exceed one per capita
\cite{pew2016}, the ownership channel is largely closed. The relevant bias
instead concerns \emph{carrier-specific representation}: the user base of any
single operator is a non-random subset of the near-universally phone-owning
population, selected by carrier preference, contract type, and voice-call
usage patterns. The direction and magnitude of this bias is an empirical
question that depends on the local market structure.

The development of methods to correct for representativeness bias in mobility metrics research has received limited attention. Although the presence of bias in mobile phone data has been recognized since the early use of Call Detail Records (CDRs), and has been extensively quantified and documented \cite{wesolowski_impact_2013, zhao_understanding_2016, Liu2024,cabrera_systematic_2025}, relatively few studies propose concrete methods to correct these biases. Existing approaches often rely on simple rescaling of observations to improve representativeness \cite{wesolowski_impact_2013, meppelink_beware_2020}, but methodological developments specifically tailored to CDR data remain limited.

More advanced approaches have been developed in related contexts. For example, some studies estimate bias and infer relative population changes over time using difference-in-differences frameworks without requiring ground truth data \cite{pestre_abcde_2020}. However, these methods are primarily designed to capture temporal dynamics rather than to correct representativeness bias in mobility metrics. Similarly, work addressing temporal bias in GPS-based mobility data has been proposed \cite{sanchez_correcting_2026}, but with a different objective.

In other domains, bias correction methods such as Multilevel Regression with Post-stratification (MRP) have been widely applied \cite{wang_forecasting_2015} and offer a promising framework for addressing representativeness issues. However, such approaches have not yet been adapted to CDR-based mobility data. Some studies have applied post-stratification techniques to adjust indicators derived from mobile phone surveys, such as total fertility rate \cite{sanchez-paez_measuring_2023}, but these approaches do not incorporate a regression component.

Recent studies on mobile phone-based mobility data continue to highlight bias as a persistent issue, emphasizing the need for new methodological developments \cite{li2024understanding, gauvin2024gaps}. This is particularly important given the increasing use of human mobility data in policy and research. Uncorrected biases may lead to misleading conclusions and suboptimal policy decisions. If mobility patterns differ systematically across the demographic groups that are differentially represented, the bias in aggregate estimates can be substantial.

In this paper, we make three contributions. First, we quantify the
representativeness bias in CDR data from a major Chilean carrier by comparing
the demographic composition of CDR users with the 2017 national census for
the Santiago Metropolitan Region. Second, we propose multilevel regression and
poststratification (MRP) as a principled statistical correction. MRP, widely
used in political science for small-area estimation from non-representative
surveys \cite{park2004, lax2009, gelman1997}, combines a multilevel model
that relates the outcome to demographic and geographic predictors with
poststratification weights derived from census population counts. Third, we
apply this framework to the radius of gyration, a standard mobility metric,
and show that MRP-corrected estimates differ meaningfully from naive CDR
averages, particularly at disaggregated levels.

\section{Data and Context}
\label{sec:data}

\subsection{Call detail records}

We use CDR data from the former Telefónica Chile mobile network (Movistar), one of the country’s three major mobile carriers, covering May to July 2016. At that time, Telefónica held approximately 31.9\% of Chile’s mobile market. Mobile penetration was about 127 subscriptions per 100 inhabitants, implying that basic mobile access was effectively universal \cite{subtel2016telecomunicaciones}. The main representativeness concern is therefore not phone ownership, but which users are captured in this carrier’s voice-call records. The original dataset contains anonymised voice-call records for users in three Chilean metropolitan regions. We restrict the analysis to the Santiago Metropolitan Region (Región Metropolitana), where census and socioeconomic data are most complete.

For each user, we know the antenna location of each call, the time, and
call metadata (placed or received, duration) as well as the socioeconomic group (GSE) and gender. We restrict the sample to
Chilean nationals with a valid GSE classification and at least one call per
day on average, following standard filtering practices to ensure reliable
location estimates \cite{pappalardo2015}. This activity threshold is
relatively aggressive and may itself introduce socioeconomic bias if call
frequency correlates with GSE; we return to this point in
Section~\ref{sec:discussion}. After filtering, the dataset contains
approximately 118,400 user-month-weekday observations.

\subsection{Construction of mobility metrics and spatial assignment for MRP}

Home locations are assigned based on the most frequent antenna during evening
hours (20:00--06:00), a standard approach in the CDR literature
\cite{vanhoof2018, pappalardo2021}. From the sequence of antenna locations, we compute the radius of gyration (ROG), i.e the characteristic distance travelled from an individual's centre of mass, measuring the spatial extent of mobility \cite{gonzalez2008}for each user in each month, separately for weekdays and weekends.

In the framework introduced here, we operate at the level of cells defined by the unique combination of distrito, GSE, gender, month, and weekday/weekend. Radius of gyration is aggregated at this level; more exaclty, we consider the mean of the log-transformed radius of gyration within each cell.

For the MRP framework, each user must be assigned to a distrito. We consider two alternative spatial assignment strategies: (i) a point-in-district approach, where each user is assigned to the distrito of their home antenna; and (ii) an area-weighted H3 hexagon approach around each antenna. As shown in Appendix~\ref{app:robust_spatial}, the area-weighted H3 redistribution yields district-level mean radius of gyration values that are virtually identical to the baseline specification, with a Pearson correlation of $r = 0.9999$ and a maximum posterior mean shift of 0.036 log-units. Overall, both user assignment strategies lead to highly consistent distributions.

\subsection{Socioeconomic data}

Socio-economic data are derived from two sources. The first is the CDR-based GSE, a label assigned by Telefónica at the individual level (originally defined at an area level but provided at the user level in the data). TThe second is the GSE distribution derived from an index called ISMT (defined below), which provides the census-level population benchmark used to correct for representativeness bias.

\paragraph{Individual GSE from CDR.} Each user in the CDR dataset is assigned a
socioeconomic group (GSE: ABC1, C2, C3, D, E) by Telefonica based on the user's salary information. This per-user attribute is included in the raw CDR extract and is
used directly as a predictor variable in the MRP model. We do not derive or
impute individual GSE from any external source.

\paragraph{Population GSE distribution from ISMT.} To construct the poststratification frame, we require the true distribution of GSE in the population by geographic unit. For this purpose, we use the \texttt{ismtchile} R package \cite{RosasAraya2023ismtchile}, which provides the Territorial Sociomaterial Index (ISMT, \textit{Índice Socio Material Territorial}), version 3, based on the 2017 census. The ISMT reports household counts by socioeconomic group at the census-zone level (\textit{redcodes}, roughly equivalent to census tracts), which allows us to derive the relative distribution of GSE within each unit rather than exact counts at the district level. We aggregate these data from redcodes to census districts (\textit{distritos}) to obtain the conditional distribution $P(\text{GSE} \mid \text{distrito})$. This distribution, expressed as proportions of individuals across GSE categories within each district, is then used in the poststratification frame alongside census population counts and gender proportions (Section~\ref{sec:poststrat})..

\subsection{Census data}

We use the 2017 Chilean Census of Population and Housing \cite{ine2017} for
two purposes: (1) to compare the demographic composition of CDR users with
the true population, and (2) to construct the poststratification frame for
MRP. The census provides person-level records with sex and geographic
identifiers (region, provincia, comuna), which we aggregate to obtain
population counts by gender and geographic unit.

\subsection{Geographic units}

Our analysis operates at three nested geographic scales within the Santiago
Metropolitan Region:
\begin{itemize}
  \item \textbf{Comuna} (47 units): Administrative divisions, the primary unit
    for the random intercept in our multilevel model.
  \item \textbf{Distrito censal} (329 units): Census districts, the fine
    geographic unit for the poststratification frame. Each distrito nests
    within exactly one comuna.
  \item \textbf{Zona censal / redcode} (730 units): Census zones from ISMT,
    the finest level at which GSE classifications are available. Each zona
    nests within exactly one distrito; we aggregate GSE proportions from
    zonas to distritos.
\end{itemize}

\section{Quantifying representativeness bias}
\label{sec:bias}

To quantify the representativeness bias, we compare the demographic
composition of CDR users with census population counts. For each redcode, we
compute the number of CDR users by GSE group and compare with the census
population in the same categories.

Table~\ref{tab:bias} presents the aggregate results for the Santiago
Metropolitan Region.

\begin{table}[H]
\centering
\caption{CDR vs.\ census composition and penetration rates by GSE group,
Santiago Metropolitan Region.}
\label{tab:bias}
\begin{tabular}{lrrr}
\toprule
GSE & CDR (\%) & Census (\%) & Penetration \\
\midrule
ABC1 & 14.4 & 31.1 & 0.022 \\
C2   & 23.1 & 15.7 & 0.069 \\
C3   & 26.7 & 25.0 & 0.050 \\
D    & 30.0 & 23.6 & 0.059 \\
E    &  5.7 &  4.6 & 0.058 \\
\bottomrule
\end{tabular}
\end{table}

Contrary to what one might expect, the highest-income group (ABC1) is the
most \emph{under}-represented in the CDR data: it accounts for 14.4\% of CDR
users but 31.1\% of the census population, yielding the lowest penetration
rate (0.022). Middle-income groups C2 and D are overrepresented, with
penetration rates of 0.069 and 0.059, respectively. The penetration rate is
not monotonically related to socioeconomic status; rather, C2 has the highest
rate, with a factor of 3.1$\times$ between the highest (C2) and lowest
(ABC1) penetration rates.

\begin{figure}[H]
\centering
\includegraphics[width=\textwidth]{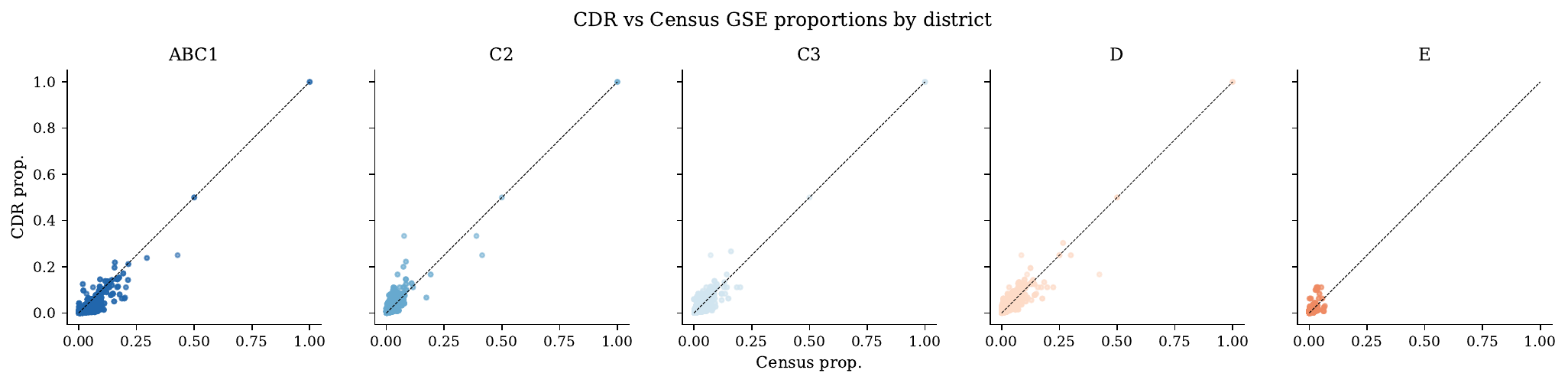}
\caption{CDR vs.\ census GSE proportions at the redcode level.}
\label{fig:scatter}
\end{figure}

Figure~\ref{fig:scatter} shows the CDR versus census GSE proportions at the
redcode level, confirming that the bias is pervasive across geographic units.

\begin{figure}[H]
\centering
\includegraphics[width=0.7\textwidth]{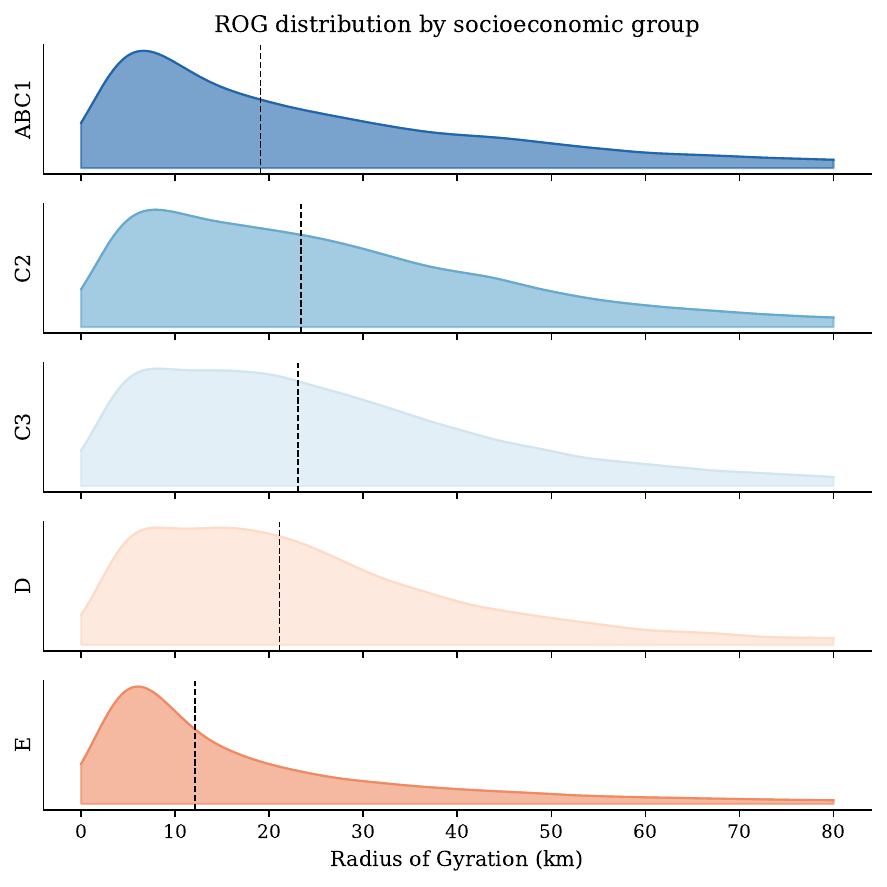}
\caption{Distribution of radius of gyration by GSE group.}
\label{fig:rog_ridge}
\end{figure}

Because mobility metrics vary with socioeconomic status (Figure~\ref{fig:rog_ridge}
shows the distribution of radius of gyration by GSE, with naive averages
decreasing from ABC1 at 27.4\,km to E at 21.5\,km), this compositional bias
translates directly into distorted aggregate mobility estimates.

\section{Multilevel Regression and Poststratification}
\label{sec:mrp}

\subsection{MRP framework}

MRP builds on earlier work on poststratification and hierarchical regression,
with its modern formulation usually traced to \cite{gelman1997} and its influential application to small-area public-opinion
estimation developed by \cite{park2004}. It was later
widely adopted in political science for estimating state-level public opinion
from national surveys \cite{lax2009, buttice2013}. To our knowledge, we are
the first to adapt this framework to correct mobility estimates from
non-representative CDR data.

\subsection{Model specification}

The model is fit on cell-level summaries: one observation per unique
combination of distrito, GSE, gender, month, and weekday/weekend. Let
$y_{ijkmt}$ denote the mean log radius of gyration for the cell defined by
distrito $i$, GSE group $j$, gender $k$, month $m$, and weekday indicator
$t$.
Our model is:
\begin{align}
  y_{ijkmt} &= \mu + \beta^{\mathrm{GSE}}_j + \beta^{\mathrm{gender}}_k
  + \beta^{\mathrm{GSE \times gender}}_{jk}
  + \beta^{\mathrm{weekday}}_t + \beta^{\mathrm{month}}_m
  + \alpha^{\mathrm{comuna}}_{c(i)}
  + \varepsilon_{ijkmt}
  \label{eq:model}
\end{align}
where:
\begin{itemize}
  \item $\mu$ is the overall intercept.
  \item $\beta^{\mathrm{GSE}}_j$ captures the main effect of socioeconomic
    group ($j \in \{\text{ABC1, C2, C3, D, E}\}$).
  \item $\beta^{\mathrm{gender}}_k$ captures the main effect of gender.
  \item $\beta^{\mathrm{GSE \times gender}}_{jk}$ captures the interaction.
  \item $\beta^{\mathrm{weekday}}_t$ and $\beta^{\mathrm{month}}_m$ control
    for temporal variation.
  \item $\alpha^{\mathrm{comuna}}_{c(i)} \sim \mathcal{N}(0,
    \sigma^2_{\mathrm{comuna}})$ is a random intercept for the comuna
    containing distrito $i$, capturing unobserved geographic heterogeneity
    across the 47 comunas of the Santiago Metropolitan Region.
  \item $\varepsilon_{ijkmt} \sim \mathcal{N}(0, \sigma^2 /
    n_{ijkmt})$ is the residual, where $n_{ijkmt}$ is the number of
    users in the cell. This observation-level weighting is equivalent
    to case-weighted regression: cells with more users contribute more
    to the likelihood in proportion to their precision.
    Appendix~\ref{app:robust_weighted} shows that results are
    substantively unchanged relative to unweighted estimation (maximum
    posterior mean shift of 0.104 log-units; all 94\% credible
    intervals overlap).
\end{itemize}

The outcome is log-transformed to accommodate the right-skewed, strictly
positive distribution of radius of gyration. The model is fit in Bambi
\cite{capretto2022}, a Python interface to PyMC \cite{salvatier2016}, using
the \texttt{numpyro} backend (JAX-accelerated NUTS) with 4 chains of 2,000
post-warmup draws each.

\subsection{Poststratification}
\label{sec:poststrat}

Let $\mathcal{J}$ index the cells of the poststratification frame, defined by
the cross-classification of distrito $\times$ GSE $\times$ gender. For each
cell $j$, we obtain the posterior predictive distribution
$\tilde{y}_j^{(s)}$ for draw $s = 1, \ldots, S$ from the fitted model
(evaluated at reference conditions: weekday, month 5). The
poststratification weights $N_j$ are the census person counts in cell $j$,
constructed under a conditional independence assumption:
\begin{equation}
  N_j = N_{\mathrm{distrito}} \times P(\mathrm{GSE} \mid \mathrm{distrito})
  \times P(\mathrm{gender} \mid \mathrm{comuna}).
  \label{eq:weights}
\end{equation}

Here $N_{\mathrm{distrito}}$ is the number of persons in the census district
from the 2017 census microdata, and $P(\mathrm{GSE} \mid \mathrm{distrito})$
is obtained by aggregating ISMT household-level GSE counts across the census
zones within each district. Because GSE is an area-level classification, the
conditional independence between GSE and gender in
equation~\eqref{eq:weights} holds by construction: the GSE composition of a
census district does not vary by gender.

The MRP-corrected estimate for any domain $D$ (e.g., a GSE group, a comuna,
or the overall population) is:
\begin{equation}
  \hat{\theta}_D^{(s)} = \frac{\sum_{j \in D} N_j \,
  \exp\!\bigl(\tilde{y}_j^{(s)}\bigr)}{\sum_{j \in D} N_j},
  \label{eq:mrp}
\end{equation}
where the $\exp(\cdot)$ back-transforms from the log scale. Uncertainty is
propagated by computing $\hat{\theta}_D^{(s)}$ for each posterior draw. The
poststratification frame contains 3,282 cells covering a total census
population of 6,217,025 persons.

\subsection{Geography-only variant}
\label{sec:geo_only}

The framework described above requires individual-level SES labels for CDR users,
as provided here by Telef\'onica's salary-based classification. When such labels are
unavailable, a reduced model retaining only geographic post-stratification can still
partially correct representativeness bias, provided that SES is spatially
clustered (a condition we verify formally before fitting the model as explained below). In that case, the geography-only multilevel model replaces the full specification in
equation~\eqref{eq:model} with:
\begin{equation}
  y_{imt} = \mu + \beta^{\mathrm{weekday}}_t + \beta^{\mathrm{month}}_m
  + \alpha^{\mathrm{comuna}}_{c(i)} + \varepsilon_{imt},
  \label{eq:model_geo}
\end{equation}
retaining the weekday and month controls and the commune random intercept, but
dropping the GSE and gender fixed effects and their interaction. The same
poststratification formula (equations~\eqref{eq:weights} and~\eqref{eq:mrp})
applies unchanged: census weights still vary by (distrito, GSE, gender); only the
model predictions (now depending solely on commune) are simplified.

To assess whether CDR coverage bias has a geographic SES structure in Santiago, we
regress the district-level log CDR/census ratio on the district's GSE composition:
\begin{equation}
  \log\!\left(\frac{n_i^{\mathrm{CDR}}}{n_i^{\mathrm{census}}}\right)
  = \gamma_0 + \sum_{j} \gamma_j\, P(\mathrm{GSE}=j \mid i) + \eta_i,
  \label{eq:ols_ses}
\end{equation}
estimated by OLS across all 329 Santiago districts (ABC1 as reference category).
The regression explains $R^2 = 0.31$ of district-level log-ratio variance
($F$-test $p = 5.7 \times 10^{-25}$), with significant coefficients for C3
($\hat{\gamma} = -4.75$, $p < 0.001$) and E ($\hat{\gamma} = -2.37$, $p = 0.007$).
Districts with a higher share of lower-income residents have systematically lower
CDR coverage relative to their census population, and this pattern is spatially
structured. Geographic post-stratification can therefore partially absorb
SES-related coverage bias.

\section{Results}
\label{sec:results}
We begin by presenting the results of the full model, followed by a comparison with a geography-only specification in Section~\ref{sec:results_geo}.
\subsection{Model diagnostics}

The full model (62 parameters: 15 fixed effects and 47 comuna-level intercepts) shows good convergence across all Markov Chain Monte Carlo (MCMC) runs: all parameters show good convergence diagnostics, with Gelman–Rubin statistics $\hat{R} \leq 1.01$, a minimum bulk effective sample size (ESS) of 782, and a minimum tail ESS of 1,534. No divergent transitions are observed.

\begin{figure}[H]
\centering
\includegraphics[width=0.8\textwidth]{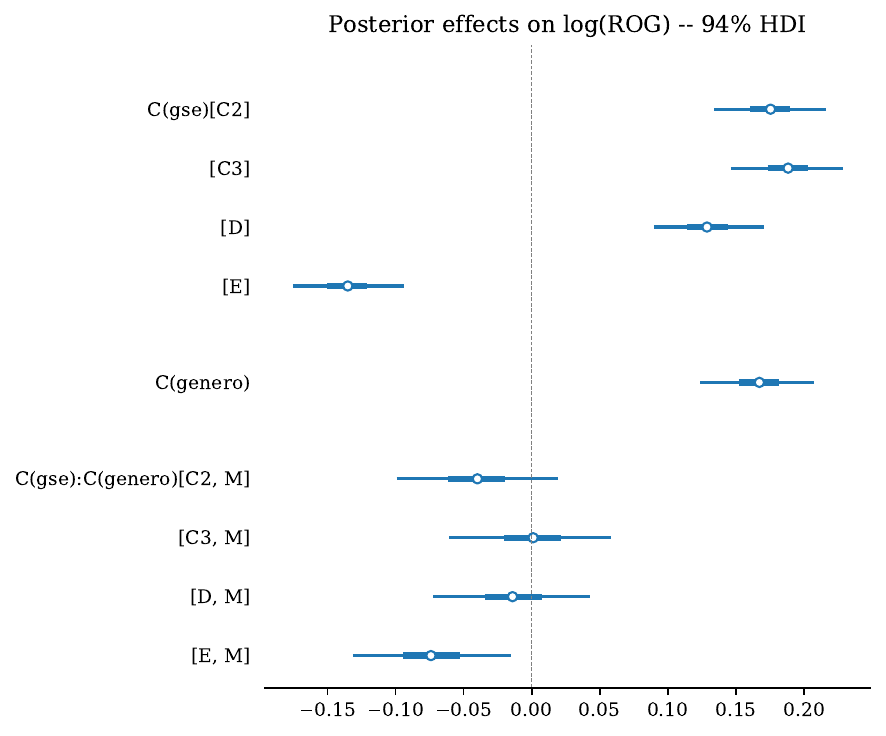}
\caption{Posterior distributions of fixed effects on log(ROG). Points show
posterior means; thick and thin bars show 50\% and 94\% highest density
intervals. GSE effects are relative to ABC1 (reference).}
\label{fig:forest}
\end{figure}

\subsection{Fixed effects}

Figure~\ref{fig:forest} shows the posterior distributions of the fixed
effects. Relative to ABC1 (the reference category), the GSE coefficients on
the log scale are: C2 $+0.175$ [0.134, 0.216], C3 $+0.188$ [0.146, 0.228],
D $+0.129$ [0.090, 0.171], and E $-0.135$ [$-0.175$, $-0.094$]. After
controlling for comuna, C2, C3, and D all have \emph{higher} conditional
ROG than ABC1; only E is substantially lower. This contrasts with the raw
(unconditional) pattern in which ABC1 has the highest naive average. The
reversal reflects geographic confounding: ABC1 zones are concentrated in
eastern Santiago comunas (e.g., Las Condes, Vitacura, Providencia) that are
relatively distant from the city centre, inflating raw ROG through longer
commutes. Once the model conditions on comuna, this spatial advantage
disappears and the within-comuna GSE gradient favours middle-income groups. Men have higher
mobility than women, consistent with prior findings \cite{gauvin2020}. The
GSE $\times$ gender interactions are small; model comparison
(Section~\ref{sec:validation}) confirms that their contribution is marginal.

\subsection{Random effects}

The comuna-level random intercept standard deviation
($\sigma_{\mathrm{comuna}}$) is estimated at 0.186 [0.148, 0.227],
indicating meaningful geographic heterogeneity in mobility beyond what is
captured by the fixed effects. As shown in the sensitivity analysis
(Section~\ref{sec:validation}), removing the comuna random intercept
substantially degrades model fit.

\subsection{MRP-corrected vs.\ naive estimates}

Table~\ref{tab:results} compares naive CDR averages with MRP-corrected
estimates. At the overall level, the MRP-corrected average ROG is
20.23\,km [19.84, 20.58], 16.6\% lower than the naive CDR average of
24.24\,km. By GSE group, the MRP-corrected gradient is C2 (21.9\,km) $>$
C3 (21.5\,km) $>$ ABC1 (20.2\,km) $>$ D (19.6\,km) $\gg$ E
(14.4\,km). Shifts from naive to corrected range from $-2.5$\,km (C3) to
$-7.2$\,km (ABC1). Notably, the MRP-corrected ranking places C2, not ABC1,
as the group with the highest spatial extent of mobility, suggesting that
middle-income groups in Santiago have the greatest true mobility once
representativeness bias is removed. By gender, women shift from 21.9 to
18.7\,km and men from 26.6 to 21.8\,km.

\begin{table}[H]
\centering
\caption{Naive CDR average vs.\ MRP-corrected radius of gyration (km) by
GSE group and overall. Credible intervals are 94\% HDI.}
\label{tab:results}
\begin{tabular}{lrrr}
\toprule
Group & Naive & MRP [94\% CI] & Shift \\
\midrule
ABC1    & 27.4 & 20.2 [19.7, 20.8] & $-$7.2 \\
C2      & 25.2 & 21.9 [21.3, 22.4] & $-$3.4 \\
C3      & 24.0 & 21.5 [20.9, 22.1] & $-$2.5 \\
D       & 23.1 & 19.6 [19.2, 20.1] & $-$3.5 \\
E       & 21.5 & 14.4 [14.0, 14.7] & $-$7.1 \\
\midrule
Overall & 24.2 & 20.2 [19.8, 20.6] & $-$4.0 \\
\bottomrule
\end{tabular}
\end{table}

Note that the naive-to-MRP comparison conflates two effects: demographic
reweighting and the back-transformation from the log scale (geometric vs.\
arithmetic mean). Both contribute to the downward shift.

\begin{figure}[H]
\centering
\includegraphics[width=0.7\textwidth]{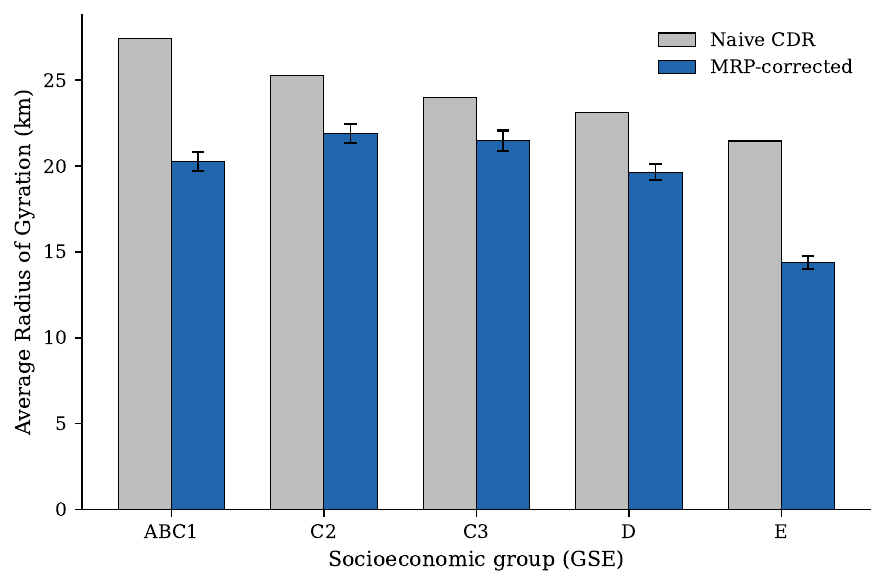}
\caption{Average radius of gyration by GSE: naive CDR average (grey)
vs.\ MRP-corrected estimate (blue, with 94\% credible intervals).}
\label{fig:mrp_gse}
\end{figure}

\begin{figure}[H]
\centering
\includegraphics[width=0.7\textwidth]{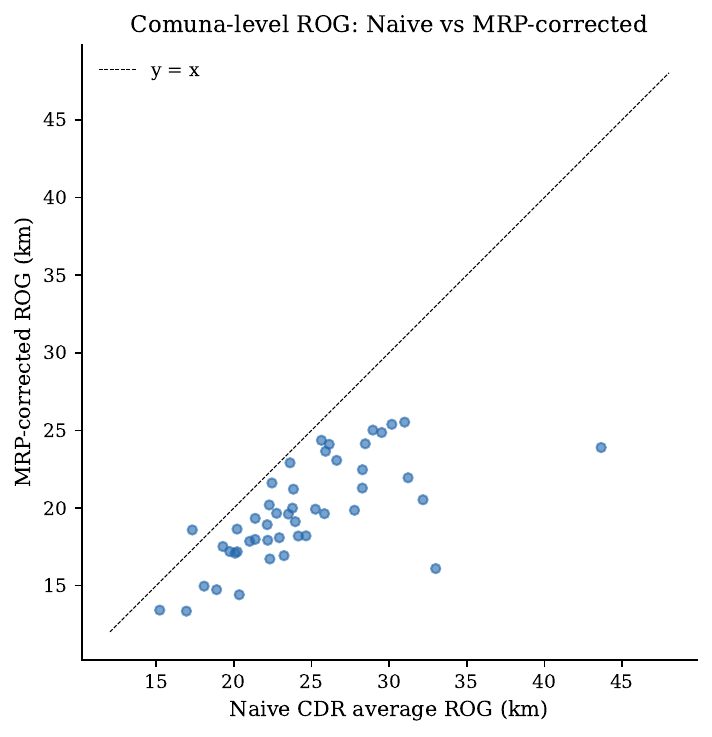}
\caption{Naive vs.\ MRP-corrected average radius of gyration by comuna.}
\label{fig:mrp_comuna}
\end{figure}

\begin{figure}[H]
\centering
\includegraphics[width=\textwidth]{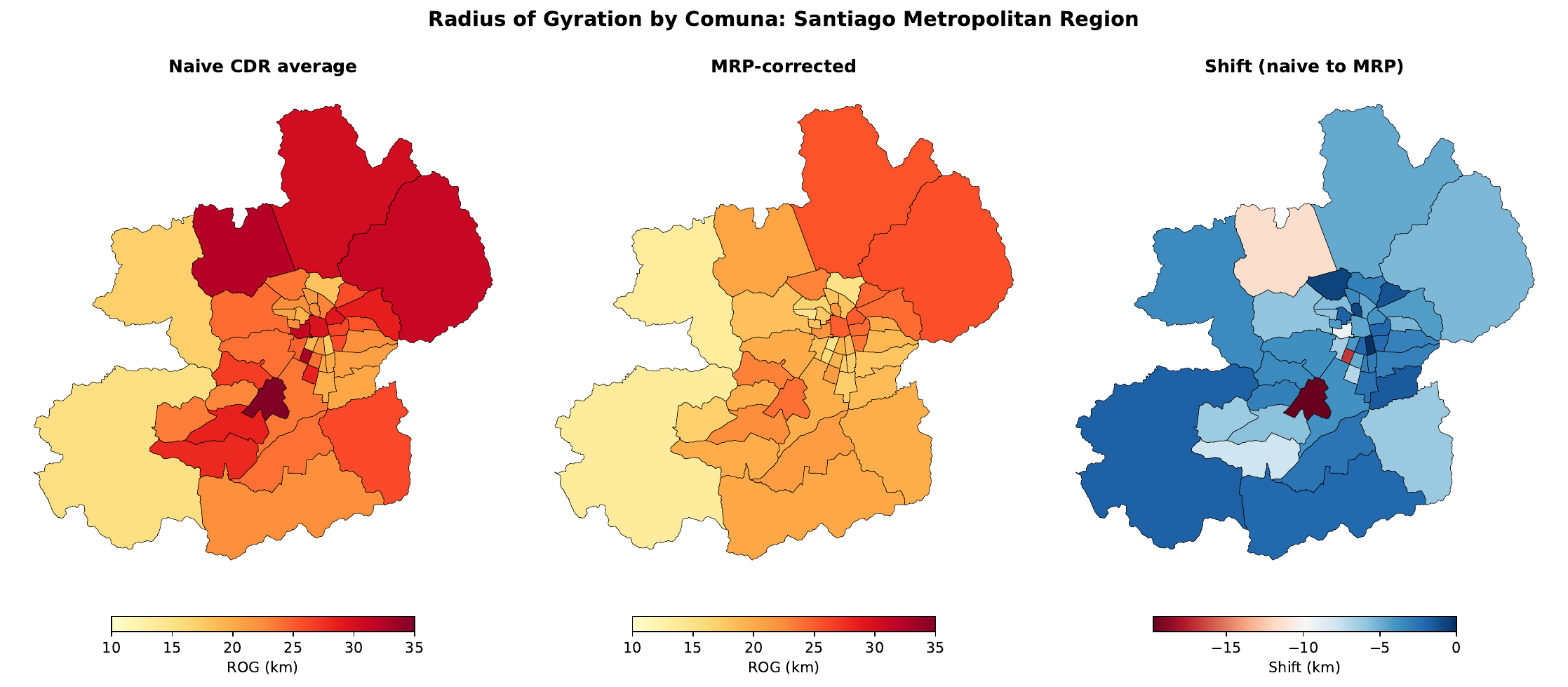}
\caption{Choropleth maps of the Santiago Metropolitan Region showing (left)
naive CDR average, (centre) MRP-corrected average, and (right) the shift from
naive to corrected radius of gyration by comuna.}
\label{fig:map}
\end{figure}

At the comuna level, the mean shift is $-4.7$\,km with a maximum absolute
shift of 19.8\,km. Comunas whose demographic composition diverges most from
the census experience the largest corrections
(Figure~\ref{fig:mrp_comuna}). The heterogeneity of the correction can be seen in the maps of Figure~\ref{fig:map}.

\subsection{Geography-only MRP}
\label{sec:results_geo}

\begin{figure}[H]
\centering
\includegraphics[width=0.85\textwidth]{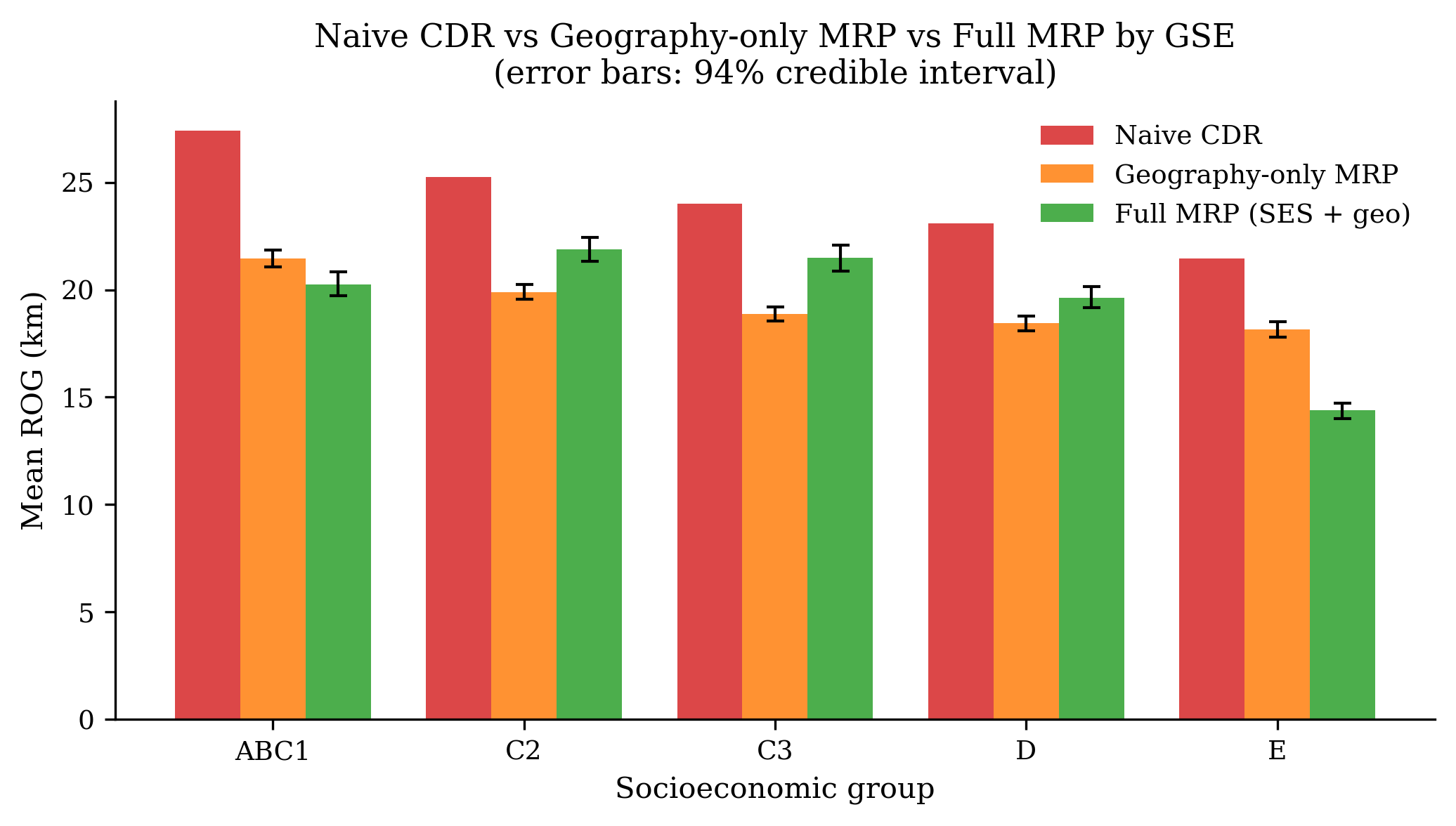}
\caption{Radius of gyration by GSE group for three estimators: naive CDR
  average (red), geography-only MRP (orange, 94\% credible intervals), and
  full MRP with SES and geography (green, 94\% credible intervals).
  Geography-only MRP corrects the aggregate bias in the right direction but
  misses the extreme deficit of the lowest-income group (GSE\,E).}
\label{fig:geo_comparison}
\end{figure}

The geography-only model (equation~\eqref{eq:model_geo}) converges well
($\hat{R}_{\max} = 1.01$, zero divergent transitions, comuna random-effect
standard deviation $\hat{\sigma}_{\mathrm{comuna}} = 0.183$, comparable to
$0.186$ for the full model). Table~\ref{tab:geo_comparison} and
Figure~\ref{fig:geo_comparison} compare the three estimators.

\begin{table}[H]
\centering
\caption{Overall radius of gyration under three estimators.
  Credible intervals are 94\% HDI.}
\label{tab:geo_comparison}
\begin{tabular}{lccc}
\hline
Estimator & Mean (km) & CI$_{3\%}$ & CI$_{97\%}$ \\
\hline
Naive CDR                   & 24.24 & --    & --    \\
Geography-only MRP          & 19.40 & 19.06 & 19.74 \\
Full MRP (SES + geography)  & 20.23 & 19.70 & 20.77 \\
\hline
\end{tabular}
\end{table}

Geography-only MRP reduces the overall estimate from 24.24\,km to 19.40\,km
[19.06, 19.74], moving in the correct direction relative to the full MRP
target of 20.23\,km. The aggregate correction (4.84\,km) is slightly larger
than the full model's (4.01\,km) because commune-level predictions do not
account for within-commune SES heterogeneity: in low-mobility communes that
house some higher-SES residents, those residents are assigned the same
uniformly low commune prediction, whereas the full model would give them a
higher one. By GSE group (Figure~\ref{fig:geo_comparison}), the geography-only
estimator undercorrects for the lowest-income group (GSE\,E: 18.16\,km
vs.\ the full MRP target of 14.39\,km, recovering only 47\% of the needed
correction) while overcorrecting for C2 and C3. Individual SES data are
therefore essential for a calibrated correction, particularly at the extremes
of the income distribution.

\section{Validation}
\label{sec:validation}

\subsection{Cross-validation}

We perform five-fold cross-validation grouped by comuna: in each fold,
20\% of comunas (and all their constituent cells) are held out, the model is
fit on the remaining 80\%, and predictions are generated for the held-out
comunas. This tests the model's ability to generalise to new geographic units,
which is the key requirement for poststratification. For a held-out comuna
with no training observations, the random intercept is drawn from the
posterior of the group-level distribution,
$\alpha_{j^*} \sim \mathcal{N}(0, \hat{\sigma}^2_{\mathrm{comuna}})$,
marginalising over the unknown group effect rather than setting it to zero.
Appendix~\ref{app:robust_cv} demonstrates this empirically on five held-out
comunas: the 94\% prediction intervals achieve 91\% empirical coverage, and
coverage at the GSE-group and comuna levels is 100\%.

Across folds, the mean RMSE on the log scale is 0.709 at the cell level, corresponding
roughly to predictions being off by a factor of $\exp(0.709) \approx 2.0$
on the original km scale. The 94\% credible interval coverage at the cell
level is 43.0\%, well below the nominal 94\%. This low coverage reflects
the fact that the model captures systematic variation (fixed effects and
random intercepts) but not the substantial within-cell individual
heterogeneity. Because MRP targets group-level aggregates rather than
individual cell predictions, coverage improves at higher levels of
aggregation (e.g., by comuna or GSE group), which is the relevant scale for
poststratification.

\subsection{Sensitivity to model specification}

We compare three model specifications using approximate leave-one-out
cross-validation (LOO-CV) via Pareto-smoothed importance sampling
\cite{vehtari2017}. LOO-CV estimates out of sample predictive performance by iteratively evaluating how well the model predicts each observation when it is left out of the fitting process. Model comparison is based on the expected log predictive density (ELPD), where higher (less negative) values indicate better predictive accuracy. We also report the standard error (SE) of the ELPD estimate and the effective number of parameters ($p_{\text{LOO}}$), which reflects model flexibility.
\begin{enumerate}
  \item \textbf{No interaction}: GSE + gender + comuna intercept
    (ELPD $= -20{,}411.8$, SE $= 132.9$, $p_{\mathrm{loo}} = 55.0$).
  \item \textbf{With interaction}: adds GSE $\times$ gender
    (ELPD $= -20{,}411.9$, SE $= 132.8$, $p_{\mathrm{loo}} = 59.1$).
  \item \textbf{No random intercept}: fixed effects only
    (ELPD $= -20{,}966.8$, SE $= 129.3$, $p_{\mathrm{loo}} = 14.8$).
\end{enumerate}

The two models with the comuna random intercept perform comparably (ELPD
difference $< 1$, well within one standard error); the interaction adds four
effective parameters for negligible improvement. The model without the random
intercept is substantially worse. Indeed, the difference in ELPD ($\approx 555$) is far larger than its associated standard error (~129), indicating decisive evidence in favor of the model with comuna random effects. We conclude
that the comuna random intercept is essential, while the GSE $\times$ gender
interaction has marginal benefit. We retain the interaction for theoretical
completeness.

\subsection{Direction of correction}

MRP-corrected estimates are lower than naive averages across all GSE groups
and 46 of 47 comunas; a single small comuna shows a marginal upward shift
of 1.3\,km. This is consistent with the joint effect of demographic
reweighting (correcting for non-uniform penetration) and the log-scale
back-transformation, which yields population-weighted geometric rather than
arithmetic means.

\section{Discussion}
\label{sec:discussion}

\subsection{When does the correction matter?}

The magnitude of the MRP correction depends on two factors: the degree of
demographic non-representativeness and the strength of the association between
demographics and the mobility outcome. In our setting, both are substantial:
CDR penetration rates vary by a factor of 3.1$\times$ across GSE groups
(C2/ABC1), and the GSE gradient in conditional mobility is strong. The
correction is most consequential at the aggregate level (16.6\% overall) and
for comunas whose demographic composition diverges most from the census
(maximum shift of 19.8\,km).

For within-group analyses (e.g., comparing weekday vs.\ weekend mobility for
ABC1 users), the correction matters less because the model conditions on the
group membership.

\subsection{Generalisability}

The MRP framework is general and can be applied to CDR data from any setting
where (1) the non-representativeness can be characterised with respect to
known demographic variables, and (2) census or survey data provide the
population distribution of those variables. The approach extends naturally to
other mobility metrics (we show results for entropy and unique locations in
the Appendix) and to other outcomes derived from CDR data.

When individual SES labels are unavailable, the geography-only variant
(Sections~\ref{sec:geo_only} and~\ref{sec:results_geo}) offers a practical
alternative: it requires only census population counts by geographic unit,
with no individual-level SES classification from the operator. In cities
with strong residential SES segregation (such as Santiago, where district
GSE composition explains 31\% of variation in district-level CDR coverage
bias), the commune random intercept absorbs much of the SES-related coverage
heterogeneity, and the aggregate correction is substantial. The residual gap
relative to full MRP is concentrated at the extremes of the income distribution,
and researchers should bear this limitation in mind when SES-disaggregated
estimates are the primary target.

\subsection{Limitations}

Several limitations should be noted. First, our analysis is restricted to the
Santiago Metropolitan Region; generalisation to other Chilean regions or
countries remains to be tested. Second, the activity filter (at least one call
per day on average) is relatively aggressive and may introduce its own
socioeconomic bias if call frequency correlates with GSE. Heavy voice callers
may differ systematically from light callers, and this selection operates
prior to our MRP correction. Third, our socioeconomic classification is at
the area level (census zone) rather than the individual level. While this is
standard in Chilean research \cite{RosasAraya2023ismtchile}, it introduces
ecological fallacy concerns. Fourth, we observe data from a single carrier
during a three-month window in 2016, so our findings reflect a temporal
snapshot. By 2016, voice calls were already in decline relative to data-based
messaging (e.g., WhatsApp), meaning our sample is further selected towards
users who still made voice calls regularly; this compounds the
representativeness problem beyond carrier market share alone. Fifth, the
naive-to-MRP comparison conflates demographic reweighting with the log-scale
back-transformation (geometric vs.\ arithmetic mean), so the reported shifts
should not be interpreted as purely demographic corrections. Finally, this study focuses on representativeness bias, but other sources of bias may also affect mobility estimates. In particular, heterogeneity in call behavior potentially correlated with socioeconomic status may influence the measurement of mobility itself, as individuals who generate more events are more likely to have their movements accurately captured. This type of bias is not addressed in the current framework.

Future work could explicitly account for this mechanism by extending the model to jointly incorporate mobility outcomes and call activity within a hierarchical framework. Another promising direction would be to leverage external data sources that are known to be unbiased, even if not fully representative, to calibrate or adjust mobility estimates, in a spirit similar to \cite{hsiao_modeling_2024}.

\section{Conclusion}
\label{sec:conclusion}

We have demonstrated that CDR data from a single carrier can have substantial
socioeconomic representativeness bias, and that this bias translates into
distorted aggregate mobility estimates. Multilevel regression and
poststratification provides a principled, well-understood correction that
leverages census data to reweight model-based predictions to the target
population.

The framework can be adopted by researchers working with any form of
non-representative passively collected data where the source of
non-representativeness is known. To facilitate adoption, we release
\texttt{mobmrp}, an open-source Python library implementing the full
pipeline (data preparation, model fitting, poststratification, and
validation) for use with arbitrary CDR and census
datasets.\footnote{\url{https://github.com/leoferres/mobmrp}} As mobile
phone data continue to inform policy decisions about urban planning,
transportation, and public health, accounting for who is and is not
observed in the data is a matter of both scientific rigor and equity.

\section*{Funding}
This research was supported by FONIS grant SA24I0124 to L.F. This work also acknowledges financial
support from the Lagrange Project of the Institute for Scientific Interchange Foundation (ISI Foundation), funded by the Fondazione Cassa
di Risparmio di Torino (Fondazione CRT).

\appendix
\section{Full model equations}
\label{app:model}

The model in Bambi formula notation:
\begin{verbatim}
log_avgROG ~ C(gse) + C(genero) + C(gse):C(genero)
           + C(is_weekday) + C(month)
           + (1 | comuna)
\end{verbatim}

Default Bambi/PyMC priors are used throughout. The model is estimated with
JAX-accelerated NUTS via the \texttt{numpyro} backend (4 chains, 2,000
post-warmup draws per chain) \cite{capretto2022, salvatier2016}.

\section{Secondary metrics}
\label{app:secondary}

We fit the same multilevel model specification for two additional mobility
metrics: Shannon entropy ($S$) and count of unique locations ($U$), both
log-transformed. All models converge well (maximum $\hat{R} = 1.00$).

For entropy, the GSE coefficients (relative to ABC1) are: C2 $-0.019$,
C3 $-0.043$, D $-0.069$, E $-0.086$; men have higher entropy than women
($+0.089$). For unique locations: C2 $-0.061$, C3 $-0.073$, D $-0.145$,
E $-0.180$; men $+0.215$. Both metrics show a monotonically decreasing
pattern from ABC1 to E, consistent with lower socioeconomic groups visiting
fewer and less diverse locations. MRP corrections shift aggregate estimates
in the same direction as for the radius of gyration.

\section{Sensitivity analysis details}
\label{app:sensitivity}

\begin{table}[H]
\centering
\caption{LOO-CV model comparison for log(ROG).}
\label{tab:loo}
\begin{tabular}{lrrr}
\toprule
Model & ELPD & SE & $p_{\mathrm{loo}}$ \\
\midrule
No interaction        & $-20{,}411.8$ & 132.9 & 55.0 \\
With interaction      & $-20{,}411.9$ & 132.8 & 59.1 \\
No random intercept   & $-20{,}966.8$ & 129.3 & 14.8 \\
\bottomrule
\end{tabular}
\end{table}

\section{Robustness checks}
\label{app:robust}

\subsection{Spatial assignment method}
\label{app:robust_spatial}

The baseline analysis assigns each user to the distrito of their home
antenna (point-in-district). A natural alternative is to map each antenna
to an H3 hexagonal cell at resolution 8 ($\approx 0.74\,\mathrm{km}^2$),
aggregate CDR values per hex cell, and redistribute them to distritos
proportionally to the fraction of each hex cell overlapping each distrito
(area-weighted redistribution).

We apply this alternative to the Santiago data. The two methods cover 451
and 424 distritos respectively; among the 420 distritos present in both,
the Pearson correlation of district-level mean ROG is $r = 0.9999$.
Refitting the full MRP model on the hex-weighted data, the maximum shift
in any fixed-effect posterior mean is 0.036 log-units (GSE group C2), and
all 94\% credible intervals overlap with the baseline. Model convergence
is identical ($\hat{R}_{\max} = 1.01$). We conclude that the results are
insensitive to the spatial assignment method.

\subsection{Population-weighted likelihood}
\label{app:robust_weighted}

Because the model is fit on cell-level averages, cells with few users
are noisier than cells with many users but are treated identically in an
unweighted likelihood. Section~\ref{sec:mrp} specifies the residual as
$\varepsilon_{ijkmt} \sim \mathcal{N}(0, \sigma^2 / n_{ijkmt})$, which
down-weights small cells. Here we verify that this choice does not drive
the results by comparing the weighted specification against an unweighted
baseline ($\varepsilon_{ijkmt} \sim \mathcal{N}(0, \sigma^2)$).

The maximum shift in any fixed-effect posterior mean between the two models
is 0.104 log-units (approximately 11\% on the ROG scale), concentrated in
the GSE main effects. The gender coefficient shifts by 0.003 log-units, and
GSE $\times$ gender interactions shift by 0.003 to 0.044 log-units. All 94\%
credible intervals overlap substantially. The direction and ordering of
all effects are unchanged. We use the weighted specification as the default
because it is the more principled treatment of aggregated data.

\subsection{Cross-validation for unseen communes}
\label{app:robust_cv}

The five-fold CV in Section~\ref{sec:validation} holds out entire comunas.
For a held-out comuna, no likelihood information updates its random
intercept during training. The unknown intercept is drawn from the posterior of the
group-level distribution:
\begin{equation}
  \alpha_{j^*} \sim \mathcal{N}(0,\, \hat{\sigma}^2_{\mathrm{comuna}}),
\end{equation}
where $\hat{\sigma}^2_{\mathrm{comuna}}$ is itself a posterior quantity
learned from the observed comunas. This marginalises over the unknown group
intercept rather than setting it to zero or using a plug-in estimate,
producing prediction intervals that are appropriately wider for unseen
comunas than for observed ones.

To illustrate this empirically, we hold out five comunas (two large: 1,678 and 1,194 cells; three small: 119, 98, and 60 cells), fit the model on the remaining 42, and generate predictions for the held-out data. We then evaluate how often the true observed values fall within the model’s 94\% prediction intervals. Across the 3,149 held-out cells, 91\% of the observed values lie within these intervals, which is close to the expected 94\%. When results are aggregated at the GSE-group and comuna levels, all observed values fall within the corresponding prediction intervals.

\end{document}